\documentclass[preprint]{aastex}

\usepackage{graphicx}

\usepackage{graphics}
\usepackage{epsfig}

\usepackage{natbib}

\newcommand{\be}{\begin{equation}}
\newcommand{\ee}{\end{equation}}
\newcommand{\nn}{\mbox{} \nonumber \\ \mbox{} }
\newcommand{\ba}{\begin{eqnarray}}
\newcommand{\ea}{\end{eqnarray}}

\newcommand{\B}{{\bf B}}

\newcommand\etal{\textit{et al.\ }}
\newcommand\eg{\textit{e.g.\ }}

\newcommand{\Bf}{{magnetic field\,}}
\newcommand{\Bfs}{{magnetic fields\,}}

\begin{document}

\title{Hartmann flow with Braginsky viscosity: a test problem for ICM plasma}
\author{Maxim Lyutikov
\\
Department of Physics, Purdue University, 525 Northwestern Avenue
West Lafayette, IN
47907-2036 }


\begin{abstract}
We consider a Hartmann layer, stationary flow of a viscose and resistive fluid between two plates with superimposed transverse magnetic field,  in the limit of  gyrotropic  plasma, when viscosity across the field is strongly suppressed. For zero cross-field viscosity, the problem is not well posed, since viscosity then vanishes on the boundaries and in the middle of the layer, where there is no longitudinal field. An additional arbitrarily small isotropic viscosity allows one to find magnetic field and velocity profiles which are independent of this viscosity floor and  different from flows with isotropic viscosity.
Velocity sharply rises in a thin boundary layer, which thickness depends both on the Hartmann number and  on the Lundquist number of the flow.

The implication of the work is that,  in simulating ICM dynamics it is imperative to use  numerical schemes which take into account anisotropic viscosity. Although magnetic fields  are dynamically subdominant in the ICM  they do  determine its the dissipative properties,  stability of embedded structures and the transition to turbulence.
\end{abstract}


\maketitle

\section{Introduction}

Gasdynamical interactions  of  magnetized flows  in the cores of clusters of galaxies play an important role in formation of the observed morphological structures, \eg\ through expansion of AGN blown bubble  into ICM medium and resulting plasma heating,  
and  interaction  of two gas components in merging clusters \cite[\eg][]{MarkevitchVikhlinin07}. 
Majority of theoretical work on these topics has been numerical, mostly using the existing fluid and MHD codes, like ZEUS. 
Unfortunately, simple hydrodynamic models "face multiple failures" \citep{Reynolds05} for cluster cores.
Perhaps, the most evident  example is  the  Raleigh-Taylor and Kelvin-Helmholtz instabilities of AGN blown bubbles, which disrupt the bubbles approximately on one rise time \citep[\eg][]{Kaiser05}.
 On the other hand, artificial fiddling with viscosity (which is usually parametrized with respect to 
 \cite{Braginsky} value; this procedure is not justified in ICM  \citep{Schekochihin05,lyutikov07}), shows that "modest" changes of shear viscosity lead to qualitatively different results \citep[\eg][ increased viscosity make the ICM plasma gel-like and quenches the instability]{Reynolds05,Sijacki}. 

One of principal reasons, perhaps,  for the  failure  of simple MHD codes
  is that they use {\it isotropic} viscosity, while   ICM  plasma is strongly gyrotropic, in a sense that it is  weakly collisional, $r_L \ll \lambda \leq L$ 
($r_L$ is ion Larmor radius, $\lambda$ is mean free path and $L$ is a typical size of the system),   and weakly magnetized, kinetic pressure $p$ is much larger than magnetic pressure, $\beta = 8 \pi p/B^2\gg 1$ ($B$ is a typical \Bf).
In ICM the  ion  Larmor radius $r_L \sim 10^{8-9}$ cm is many orders of magnitude smaller than the  mean free path $\lambda \sim 10^{22-23} $ cm and the size $L\sim 10^{24}$ cm, while $\beta \sim 100$ \citep{ct02}.

In a strongly gyrotropic plasma the 
 local  transport properties, primarily viscosity and conductivity,  become anisotropic \citep{Braginsky}. The effects of anisotropic viscosity and conductivity are expected to change substantially  the results of ICM simulations. As a simple example, note that \Bf\ drapes around the contact surface separating the two interacting media  \citep{lyutikov07}. In a strongly gyrotropic plasma the shear viscosity inside the draping layer, with a  flow along \Bf\ lines, then becomes 0. This runs contrary to the idea that high viscosity may provide stabilization  \cite{Reynolds05}. (Note that draping itself can provide stabilization against KH instability, in a way similar to the effect of a thin oil layer on water waves \citep{Dursi07}).
 
  Understanding the basic properties of  strongly gyrotropic plasma is imperative for further progress, especially for parametrization of the 'sub-grid' physics in large numerical simulations.  In this Letter we adapt one of the basic solutions of MHD, a Hartmann flow  \citep[\eg][]{LLVIII}, to the case of anisotropic viscosity.

\section{Hartmann flow with anisotropic viscosity}
 Consider a one-dimensional  (along $x$ direction)   flow  of a weakly collisional  plasma between two plates located at $z= \pm a$  with a superimposed external \Bf (generally, oblique). This is meant to represent a boundary layer during interaction of two plasma flows in the  ICM.
    In the Chew-Goldberger-Low (zero Larmor radius and neglecting heat fluxes) approximation \citep{Chew} the  equations of  resistive plasma flow read \citep{Kulsrud}:
\ba &&
{d { \bf v} \over dt} = - \nabla \left( P_\perp + B^2/2 \right)
 +  \nabla \left(\hat{b} \hat{b}(P_\parallel -P_\perp - B^2)\right)
\nn &&
P_\parallel -P_\perp= 3\eta_0 d_t \ln B= 3 \eta_0  \left(\hat{b} \hat{b} \nabla  { \bf v} \right) 
\nn &&
\partial_t \B={\rm  curl }  { \bf v} \times \B + \eta_r \cdot  \Delta \B
\ea
where $\eta_0$ is the first Braginsky coefficient \citep{Braginsky},  resistivity  $\eta_r$ is a tensor and $\hat{b}=\B/B$ is a unit vector along \Bf. We also absorbed a factor $\sqrt{4 \pi}$ into definition of \Bf,

Assuming that all quantities are independent of $x$ and $y$,
from div$ \B =0$ we find $B_z = {\rm const } = B_0$. Introducing $B_x= B_0 f(z) $, $v_x=v$,  the $x$ component of  Euler equation gives
\be
B_0^2 f' +  3 \eta_0 \partial_z \left(  { v^{\prime } f^2 \over (1+f^2)^2} \right) - 
\partial_x P_\perp =0
\label{eq1}
\ee
$ \partial_x P_\perp= \Delta P/L $ is  constant pressure gradient driving the flow, $\Delta P$ is a drop in pressure over the length $L$.  Eq (\ref{eq1}) can be integrated once:
\be 
B_0^2 f - {\Delta P \over L }  (z-z_0) +  3 \eta_0   { v{\prime } f^2 \over (1+f^2)^2} =0
\label{eq2}
\ee
Integration constant  $z_0$ is a point where $f=0$.

Assuming that the $y$ component of  \Bf\ vanishes, so that current flows across \Bf\ ${\bf j} = j_y {\bf e}_y$, 
the
resistivity equation gives:
\be
v'
= - \eta_\perp f^{\prime \prime}
\label{v}
\ee
Here $\eta_\perp$ is resistivity across \Bf.
Finally, eliminating $v'$ from Eq. (\ref{eq2}) using Eq. (\ref{v}),  we get equation for $f$:
\be
B_0^2 f - {\Delta P \over L }  (z-z_0) -{   3 \eta_0  \eta_\perp}    {f^{\prime \prime}  f^2 \over (1+f^2)^2} =0
\ee
This is the main equation which  determines the structure of the flow. In case of anisotropic viscosity it's a nonlinear equation.   It  differs from the case of isotropic viscosity  \citep[\eg][\S 67]{LLVIII} by  a different viscosity term:
\be 
\eta_{eff} =  3 \eta_0 { f^2 \over (1+f^2)^2}
\ee
Renormalizing, $ 
z \rightarrow z a, \, z_0 \rightarrow z_0 a$ and introducing 
$\alpha= (B_0^2/\Delta P) (L/a)$, 
we find
\be
z_0-z  + \alpha \left(  f - {1\over G^2} {f^{\prime \prime}  f^2 \over (1+f^2)^2}  \right) =0
\label{main}
\ee
where 
\be
G = { B_0 a \over \sqrt{ 3 \eta_0 \eta_r} }
\ee
we identify as   Hartmann number.


Boundary conditions are somewhat tricky in our case. From the continuity of the tangential \Bf (assuming there is no surface current) and from the symmetry of the flow,  it is required that
$f=f_0$ at $z= \pm a$  and $f=0$ at the  point $z_0$, where $f_0$ is an imposed longitudinal \Bf. Below we consider a case when there is no superimposed longitudinal magnetic field,   $f_0=0$ and   $z_0=0$.
 
 On the other hand, at surfaces where $B_x \propto f=0$ (at the boundaries and in the middle) there is no viscosity, see Eq. (\ref{main}),  so the usual conditions of $v=0$ at $z= \pm a$  need not to be satisfied. Thus, in principle,  the flow may  slip along the boundaries and have a discontinuity in the middle. This will make the problem unsolvable as the order of the ODE would be higher than the number of boundary conditions. 

In fact, it is necessary to assume some floor for  viscosity  to get   a 
physically meaningful solution of  Eq. (\ref{main}) even if we just impose condition $f=0$ at $z= \pm a$.
 Near points $f=0$, Eq. (\ref{main}) reduces to
$ f=  (z-z_0)/  \alpha $. So that  derivative  of $f$ at these points has a  definite sign, given by the parameter $\alpha$. Thus, $f$ cannot be zero more than once. Since $f=0$ in middle of the layer, this would  clearly violate  the 
conditions  that parallel
 \Bf vanishes on the walls.
 
\begin{figure}
\includegraphics[width=\linewidth]{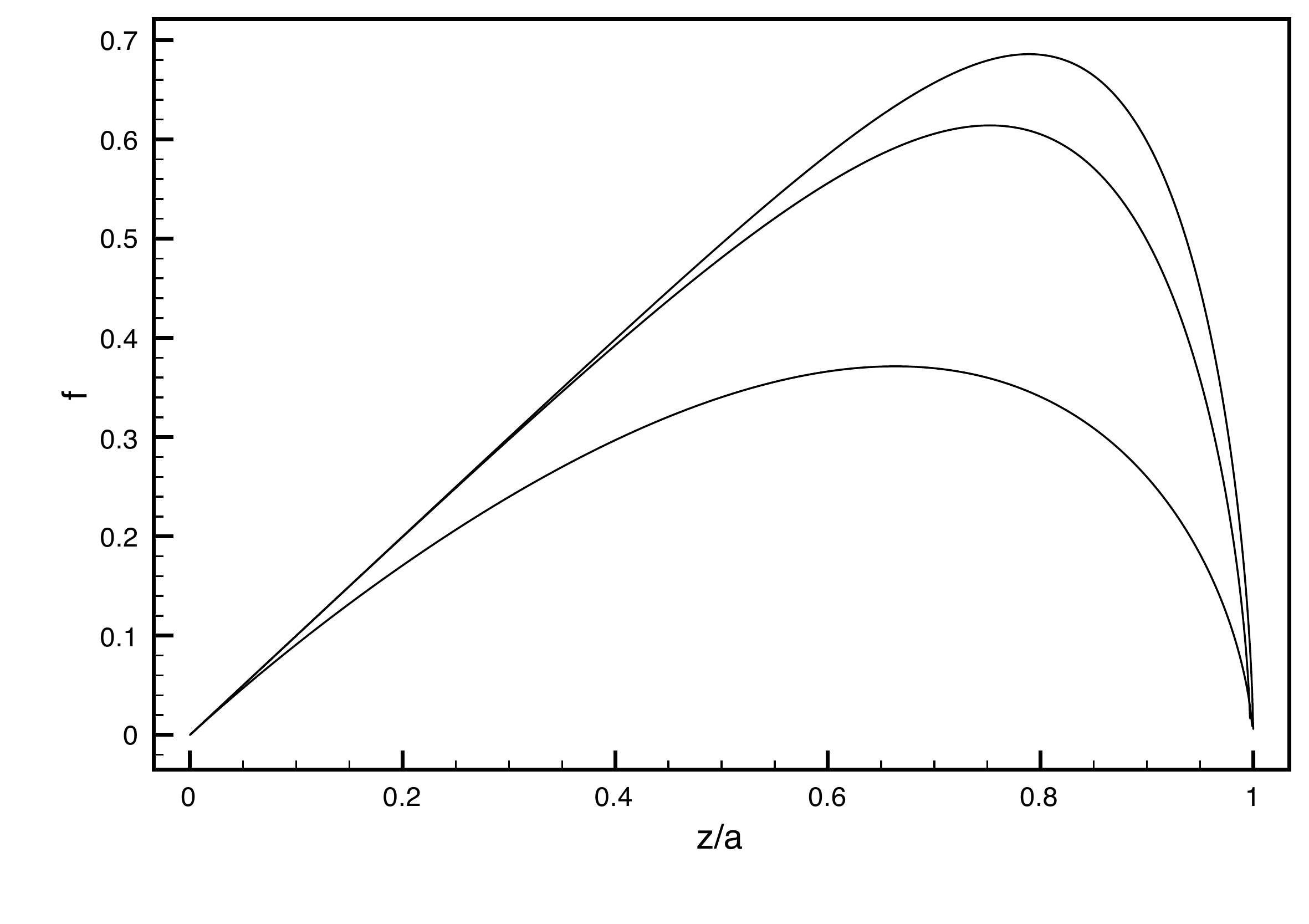}
\caption{Function $f$ (longitudinal magnetic field) for $\alpha=1$ and different $G=1, \, 3.1, \, 4.5$ (bottom to top)  (larger  values of $G $ are difficult to resolve numerically). }
\label{Fig1}
\end{figure}

\begin{figure}[h]
\includegraphics[width=\linewidth]{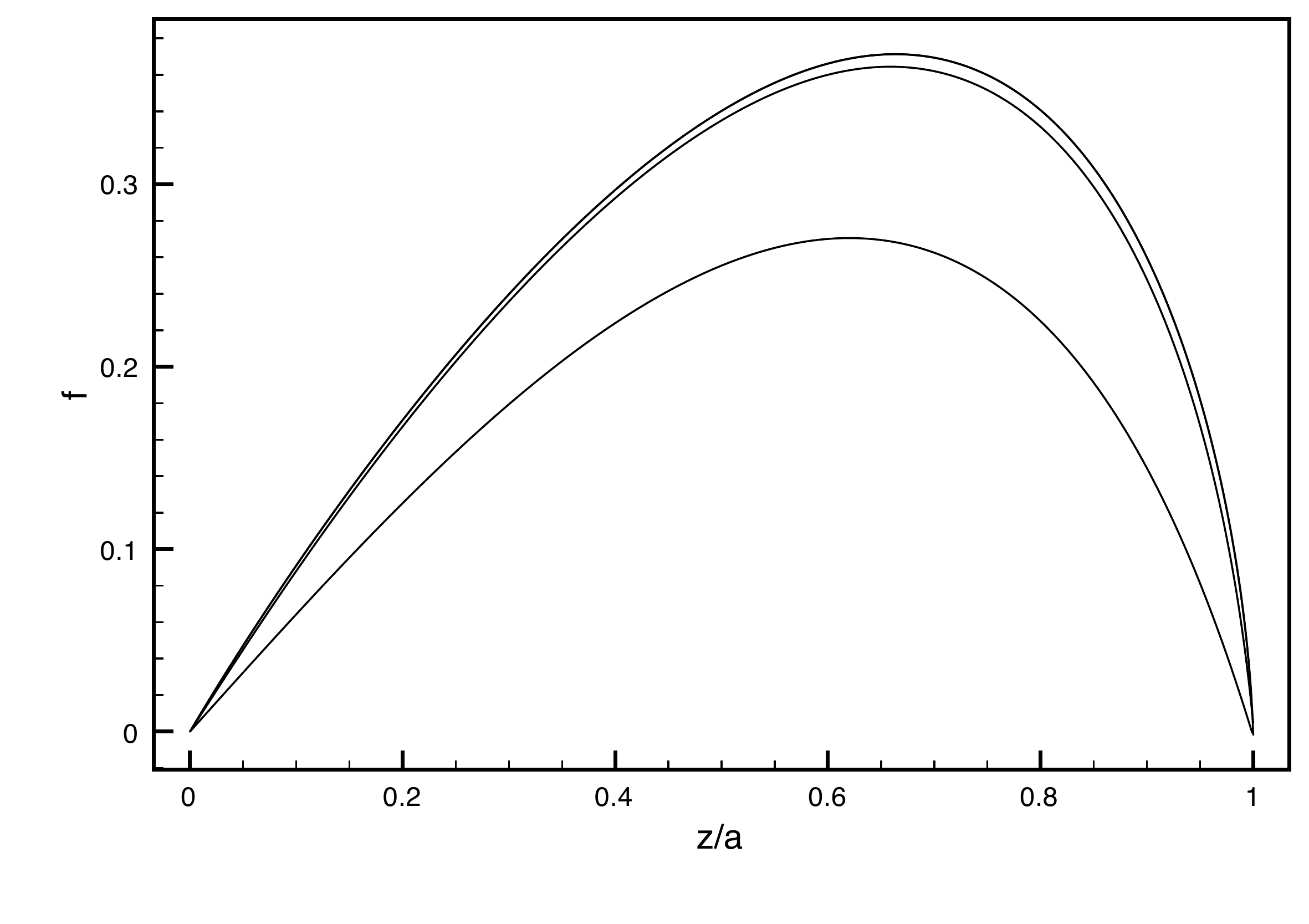}
\caption{Effect of finite isotropic viscosity on the structure  of \Bf for $f_r=0.1, 0.005, 0.0001, 10^{-10}$  (bottom to top), $\alpha = G =1$. For small enough $f_r$ solutions are nearly independent of the exact value of  $f_r$.
}
\label{comparefr}
\end{figure}

Introducing resistivity floor $f_r$ in Eq. (\ref{main}) by substitution $f^2 \rightarrow f^2 + f_r$, we can
integrate  Eq.  (\ref{main})  numerically, see Fig.(\ref{Fig1}).
 For sufficiently small values of this viscosity floor the final result is independent of its exact value, Fig. (\ref{comparefr}).

The somewhat unphysical value of $\Delta P \over L$ can be expressed in terms of bulk velocity in the middle of the layer $v_0$:
\be
{\Delta P \over L} \sim {v_0 B_0^2 \over G^2 \eta_\perp}={v_0 \eta_0 \over a^2}
\ee
This gives $\alpha= G^2 {\eta_\perp \over v_0 a}= G^2/L_u$, where we identified the ratio $v_0 a/  \eta_\perp$ with magnetic Reynolds, or Lundquist number $L_u$. Parameter $\alpha$ plays an important role in determining dynamics of the layer. If viscosity is dominated by ion-ion collisions, we may estimate 
$\alpha \sim 1/( \beta M_s K_n)$, where $\beta $ is plasma beta parameter (ratio of kinetic to magnetic pressure), 
$M=v_0/c_s$ is the flow Mach number (ratio of velocity to sound speed $c_s$),  $K_n=\lambda/a$ is  
Knudsen number, the  ratio of the mean free path $\lambda$ to the characteristic length scale $a$.

These solutions are quite different from the case of isotropic viscosity  \citep{LLVIII} and have a number of particular feature. On the one hand, 
for large $G \gg 1$, in the bulk of the flow the profile of $f$ is linearly increasing, $f \sim z/\alpha$ (flat velocity profile),  similar to the isotropic viscosity case. The parallel \Bf drops  back to 0 within a narrow boundary layer.
Let us estimate the thickness of this boundary layer.  Near the boundary $f \rightarrow 0$, so that an approximate solution of  Eq. (\ref{main}) is 
\be 
f={z\over \alpha} - {a\over \alpha} \left( {z\over a}\right)^{(1+ \sqrt{1+ 4 G^2 \alpha^2})/2}
\label{f}
\ee
If we define the thickness of the boundary layer $\delta$ when $f'=0$, we find
\be
{\delta  \over a} \sim   {\ln (\alpha G) \over \alpha G}  \sim \ln (\alpha G)  { L_u \over G^3}
\label{delta}
\ee
where we assumed $G\alpha \gg 1$.
This expression can be compared with the case of isotropic viscosity, when $\delta_i /a \sim 1/G$. The ratio
$\delta/\delta_i \sim 1/\alpha = L_u/G^2$. Thus, parameter $\alpha= G^2/L_u$ measures the relative concentration of a \Bf profile toward the wall; for $\alpha > 1$  the boundary layer in case of  anisotropic viscosity is narrower than in the isotropic case. 

In most applications the ratio $L_u/G^3$ is very small. For example, for the typical parameters of intercluster medium, velocity $\sim 1000 $ km/s and layer thickness $\sim 1 kpc$,  plasma beta $\beta =100$, we estimate
$L_u = 10^{27}$, $G= 10^{12}$, so that 
   $L_u/G^3 \sim 10^{-9}$, thus  $\delta \ll a$.
\begin{figure}
\includegraphics[width=.9\linewidth]{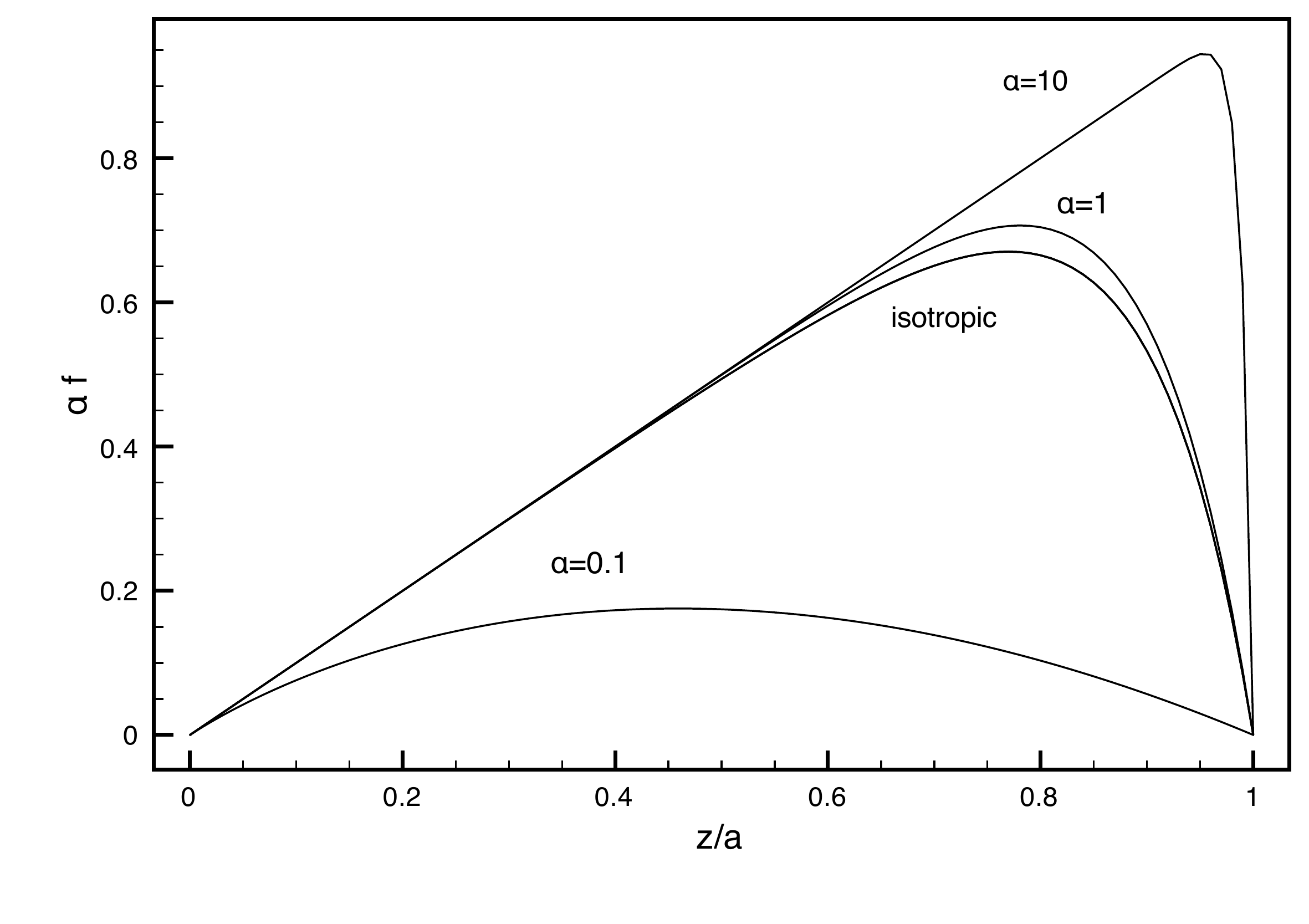}
\caption{Comparison of  analytical \Bf profiles (multiplied by $\alpha$ for better graphical representation) for isotropic \citep[][$f\propto \left( (z/a) \sinh G - \sinh(z G/a)\right)/(\cosh G -1)$]{LLVIII} and  approximate anisotropic (Eq. (\ref{f})) viscosity for  $G=10$ and different parameters $\alpha =0.1,\,1\,10$. For $\alpha > 1$, the boundary layer in case of  anisotropic viscosity is narrower than in the isotropic case. }
\label{compare}
\end{figure}



\section{Conclusion}

In this letter we considered a basic plasma physics problem: Hartmann flow with anisotropic viscosity. 
We first argued that in the case when transverse viscosity  is suppressed completely, the problem cannot be formulated in  a physically meaningful way: there should be some small isotropic contribution to viscosity.
We derived  \Bf and velocity profiles, which in the limit of small  isotropic viscosity floor  are independent of its exact value. 
These profiles are considerably different from the case of isotropic viscosity. The velocity gradients are much more concentrated close to the walls of the channel than in the case of isotropic viscosity. 

How important is the structure of a boundary layer for the overall structure of the flow? On the one hand, in a laminar regime at low Reynolds numbers, the structure of the boundary layer is probably not important: in the boundary layer the relative velocity just drops to zero, according to some law,  without affecting the overall structure of the flow. On the other hand, properties of the  boundary layer  determine its stability  and, thus, transition to turbulence \citep{LLVI}. In ICM plasma the Reynolds numbers are in the range $Re \sim 10-1000$, while, typically, transition to turbulence occurs at $Re \sim 100-1000$.  (Transition to turbulence occurs in the boundary layer). Thus, we expect that effects of anisotropic viscosity  are likely to be very important for ICM plasma, especially for determining its transition to turbulence. 

Thus, in simulating ICM dynamics it is imperative to use appropriate numerical schemes which take into account anisotropic viscosity (and to a lesser extent, conductivity), like the ones that have been applied to accretion disks \citep{sharma07}. Though \Bfs are dynamically subdominant in ICM (plasma beta  parameter is large), they do, in fact,  determine its the dissipative properties,  stability of embedded structures and transition to turbulence.
 

\end{document}